\documentclass[preprint,onecolumn,superscriptaddress,showpacs,citeautoscript]{revtex4-1}

\usepackage{amsmath,amssymb}
\usepackage{amsfonts}
\usepackage{subfigure}
\usepackage{bm}
\usepackage{graphicx}
\usepackage{appendix}
\usepackage{multirow}
\usepackage[colorlinks,linkcolor=black,urlcolor=blue,citecolor=black]{hyperref}
\usepackage{url}
\usepackage{titlesec}
\titleformat{\subsection}{\center\small\bfseries}{}{0pt}{}
\begin{document}
\title{\textbf{Practical phase-modulation stabilization in quantum key distribution via machine learning}}
\author{Jing-Yang Liu}
\author{Hua-Jian Ding}
\author{Chun-Mei Zhang}
\author{Shi-Peng Xie}
\email{xie@njupt.edu.cn}
\author{Qin Wang}
\email{qinw@njupt.edu.cn}
\affiliation{Institute of quantum information and technology, Nanjing University of Posts and Telecommunications, Nanjing 210003, China}
\affiliation{“Broadband Wireless Communication and Sensor	Network Technology”Key Lab of Ministry of Education, Ministry of Education, Nanjing 210003, China}
\affiliation{“Telecommunication and Networks”National Engineering Research Center, NUPT, Nanjing 210003, China}

\begin{abstract}
	In practical implementation of quantum key distributions (QKD), it requires efficient, real-time feedback control to maintain system stability when facing disturbance from either external environment or imperfect internal components. Usually, a "scanning-and-transmitting" program is adopted to compensate physical parameter variations of devices, which can provide accurate compensation but may cost plenty of time in stopping and calibrating processes, resulting in reduced efficiency in key transmission. Here we for the first propose to employ a well known machine learning model, i.e., the Long Short-Term Memory Network (LSTM), to predict those physical parameter variations in advance and actively perform real-time control on corresponding QKD devices. Experimentally, we take the phase-coding scheme as an example and run the LSTM model based QKD system for more than 10 days. Experimental results show that we can keep the same level of quantum-bit error rate as the traditional "scanning-and-transmitting" program by employing our new machine learning method, but dramatically reducing the scanning time and resulting in significantly enhanced key transmission efficiency. Furthermore, our present machine learning model should also be applicable to any other QKD systems using any coding scheme or QKD protocols, and thus seems a very promising candidate in large-scale application of quantum communication network in the near future.	
\end{abstract}

\maketitle

\section{INTRODUCTION}
	Quantum key distributions (QKD) \cite{BB84,Diffie,E91,PQC} can provide information-theoretic secure keys between two communicated parties, usually called Alice and Bob, and has attracted extensive attention from the scientific world since the first BB84 protocol was proposed \cite{BB84}.To date, significant progresses have been achieved in this field both theoretically and experimentally, making it moving from laboratory research to practical implementation and from point-to-point communication to multiuser complex networks \cite{WXB2005,Lo}. 
	
	In practical implementation of QKD, in order to maintain the stability and reliability of the system, especially for fast-speed QKD networks, real-time control are highly demanded due to very complex realistic environment and imperfect internal devices. In the history, the Faraday-Michelson Interferometer (FMI) was designed for phase encoding QKD systems to solve the problem of interference instability existing in Mach-Zehnder Interferometers (MZI) \cite{FM} caused by polarization changes in transmission fibers. Both theory and experiment have proven the polarization stability of FMI based QKD systems \cite{ZFHan}. Nevertheless, neither MZI nor FMI can maintain phase stability for a long time due to ineluctable internal phase shift. Therefore, it is quite common to use a "scanning-and-transmitting" program to calibrate those physical parameters in present QKD systems \cite{Dstate,ZPRAplied,chaldis,gigBB84}. Although present calibration programs can keep the stability of QKD systems, it is at the cost of transmission efficiency, since it takes time to run the calibration program and no signal data can be transmitted during the calibrating process. Traditionally, the operation mode of a "scanning-and-transmitting" program can be described by the concept of "duty ratio" \cite{phacomp} which refers to the ratio of the transmission time to the total time. Obviously, the transmission efficiency of a QKD system is closely related to its "duty ratio" and it is crucial to increase the "duty ratio" of complex QKD networks towards practicality.
	
	The Long Short-Term Memory network (LSTM) \cite{LSTM1,LSTM2} is a special kind of recurrent neural network, whose chain-like nature reveals that it is the natural architecture of neural network to use for time series \cite{timeSer1}. By virtue of these innate advantages, here we for the first time propose to employ LSTM networks to improve the "duty ratio" of present QKD systems and solve those existing problems. Furthermore, we carry out corresponding experimental demonstrations, where we take a FMI based BB84 QKD system as an example and do comparisons between adopting our present method and applying conventional "scanning-and-transmitting" model.
	
	This paper is arranged as follows: In Sec. II, we describe in detail how we construct a LSTM model to predict the variations of phase voltage and run it on a FMI based BB84 QKD system. In Sec. III, we analyze the experimental data and do comparisons between our present work and conventional work. Finally, summaries and outlooks are given out in Sec. IV.
	
\section{METHODS}

In this section, we will give descriptions on the methods including how to construct the LSTM model, how to predict the variations of phase voltages and how to carry out corresponding experimental demonstrations. Below let us introduce it step by step.

\subsection{A.\quad Establish data structure}
    
    Considering the memory characteristics of the LSTM, we divide the training data into into many sequences according to time, and at each time the data point consists of "features" and "labels". Now the purpose of applying machine learning is to construct certain mapping from the characteristics of "features" to "labels" by learning from known data.
    
    As we know, phase shift inevitably exists in phase-coding QKD systems due to variations of many factors in ambient environments, including the temperature, the humidity and the voltages. Moreover, the intensity fluctuation of a laser also has a great impact on the applied voltage of a phase modulator. Based on the above, we extract useful "features" from existing data, i.e., the operating temperature, the humidity, the intensity of a laser, and the voltages, and designate the voltage of the next moment as "labels". It is worth mentioning that, here we add five time-series displacement voltages into "features" to enhance temporal relevance of different points in time, facilitating learning. Corresponding data structure diagram is shown in Fig. 1.  
	
\begin{figure}[!htbp]
		\centering
		\includegraphics[width=0.9\linewidth]{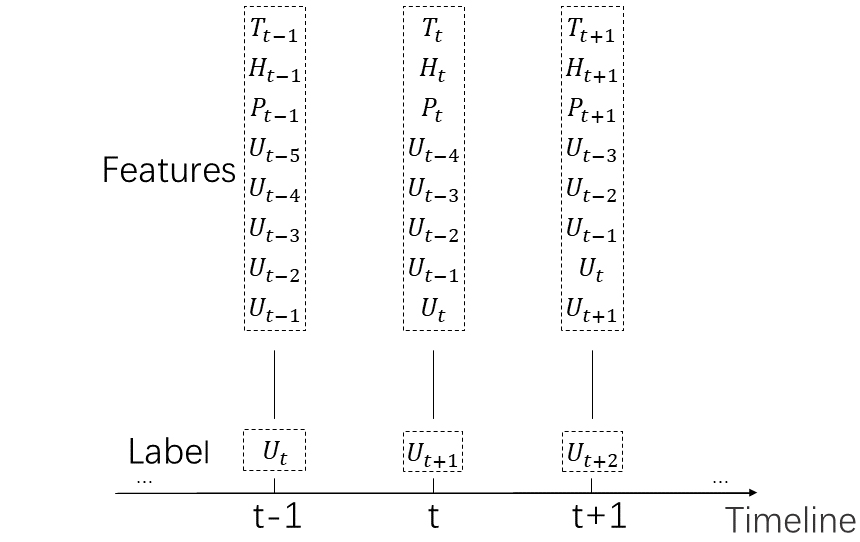}
		\caption{The data structure diagram. T, temperature; H, humidity; P, intensity of laser; U, voltage. For example, if the current time is t, then the feature of t moment is  $[T_{t},H_{t},P_{t},U_{t-4},U_{t-3},U_{t-2},U_{t-1},U_{t}]^{T}$, the label of this moment is $U_{t+1}$. }
		\label{fig:eta2m2x}
\end{figure}
	
	    Here we are doing supervised learning with the LSTM model to predict the voltage output of the traditional "scanning-and-transmitting" program. We depict the curve of zero-phase voltage versus time in Fig. 2 to indicate the variation of the zero-phase voltages over a relative long time period.
	
	\begin{figure}[htbp]
		\centering
		\includegraphics[width=0.9\linewidth]{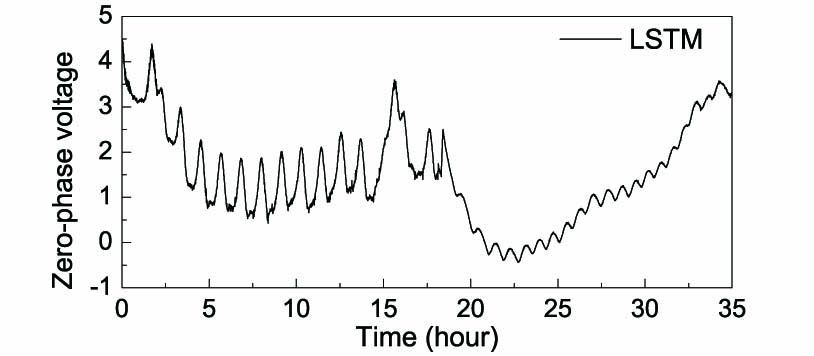}
		\caption{Zero-phase voltage versus time curve.}
	\end{figure}

\subsection{B.\quad Predict phase voltage by LSTM}	
	Before addressing our machine learning work, let us briefly review the working principles of traditional "scanning-and-transmitting" programs. Usually, Alice fixes her phase voltage, and Bob scans his phase modulation voltage across the full scale range of the Data Acquisition Chassis with a large number of steps. At each step, Bob records the photon counts through single-photon detectors (SPD), and he is able to ascertain the zero-phase voltage of the minimum count by fitting or other means, which corresponds to the phase where the destructive interference exactly takes place. According to the fitted zero-phase voltage, Alice and Bob process feedback control and key transmission. 
	
	In contrast to the above traditional program, in our present work the zero-phase voltages can be actively predicted through implementing the LSTM model instead of scanning and fitting the  interference fringes. Of course, before predicting, a large number of data points should be collected for training the LSTM network, which can be performed before real key transmission. We prepared the training data by running the FMI based BB84 QKD system with traditional "scanning-and-transmitting" program, including the scanned zero-voltages, the operating temperatures, the humidities, and the intensities of the laser. During the above "scanning-and-transmitting" process, the duty ratio of the QKD system is 0.5. The whole training data consists of 10 sets, and each set  contains a time series of 3600 data points, temporally corresponding to 20 hours.  
\begin{figure}[!htbp]
	\centering
	\includegraphics[width=1.0\linewidth]{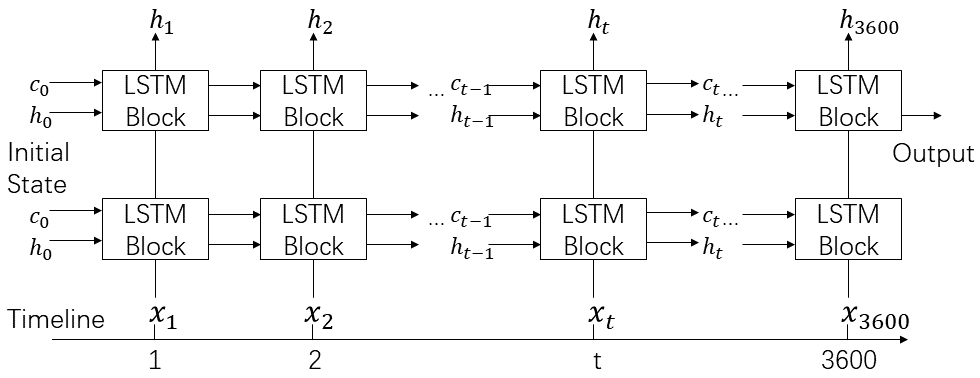}
	\caption{Diagram of a two-layer LSTM network, the length of this network is the time span of a training set. $ h_{t} $ denotes the output of last moment, $ c_{t} $ represents current cell state, $ x_{t} $ refers current input. }
	\label{fig:eta2m2x}
\end{figure}	
\begin{figure}[!htbp]
		\centering
		\includegraphics[width=1.0\linewidth]{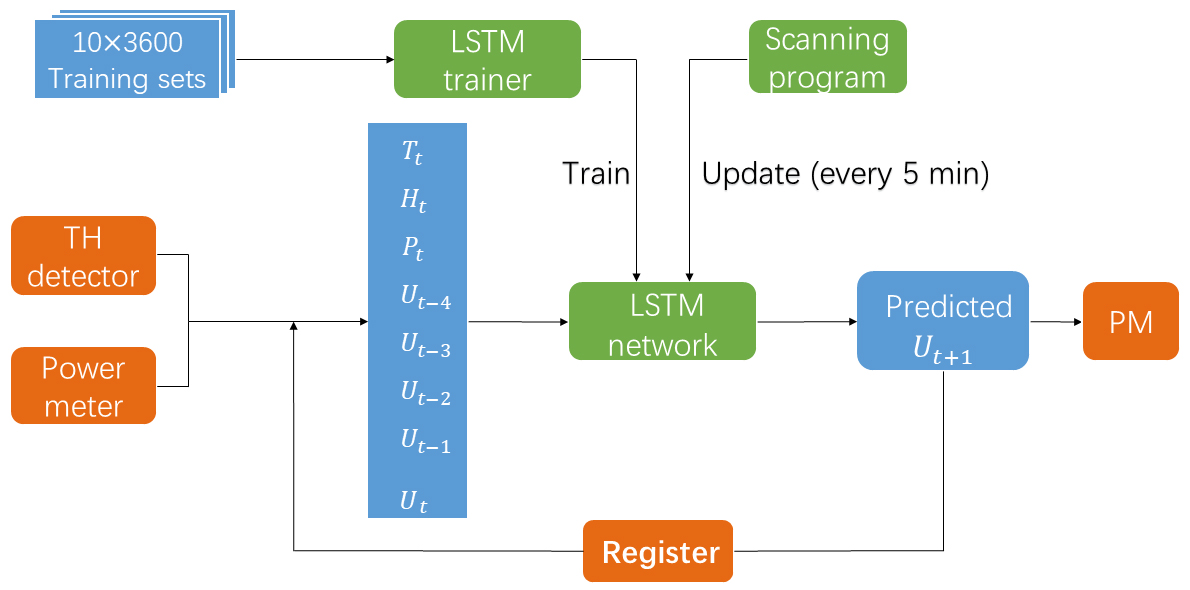}
		\caption{Data flow of predicting with a two-layer LSTM network. The round boxes represent programs and squared boxes represent data. The network first executes training with 10$\times$3600 sets of training data. After training, it can predict the zero-voltage at the next moment by reading-in current values of "features", i.e., the temperature and humidity, the optical power, and the displacement voltages predicted at the previous five moments as well. Finally, the predicted voltage is applied on the phase modulator on Bob's side for the next moment. }
		\label{fig:eta2m2x}
\end{figure}
	
	In order to enable the network executing deep learning, here we design an improved two-layer LSTM network structure and utilize a mean squared error cost function. The number of hidden neurons in the first layer is 20 and the number of hidden neurons in the second layer is 10. Moreover, an method of Z-score normalization has been adopted for all data before input into the network. The developed LSTM network structure is shown in Fig. 3. Furthermore, to ensure the long-term reliability of network  forecasting, we adopt a structure combining forecasting with updating. Specifically, we use the scanning program as a tool to update the LSTM network, eliminating the cumulative error in the  prediction process and keeping the quantum-bit error rate (QBER) of the QKD system within an acceptable level. For example, after each prediction of 25 points (250 seconds), we run the scanning program for 5 times (50 seconds), recording the zero-voltages applied on the phase modulator. Meantime, corresponding values of the ambient temperature, the humidity, the laser intensity are also returned to the network. Then the weights and the biases of the network can be updated accordingly.
	
	Although the LSTM network requires that the training data and subsequent testing data must be a continuous time series, considering the practicability, it is not possible to retrieve training data with the original QKD system every time before using the network. Therefore, we discover a way by updating the LSTM network continuously for 10 minutes every time before transmission. When the training process is good enough, it does not matter even if the training data and testing data are not continuous, because the rule of data changes has been embedded in all the weights and biases of the network, we only need to fine-tune the network parameters to adapt to the current time series through a period of updating. 
	
	We use Adam as the optimization algorithm for 270 epochs, which takes approximately 15 minutes on our PC (CPU:Intel Core i7 9700@ 3.6GHz; GPU: NVIDIA GeForce RTX 2080; RAM: DDR4 8GBytes). The initial learning rate is 0.02, and drop 80\% every 100 epochs. Batch size is 3600 data points. 

\subsection{C.\quad Experiment}	

\begin{figure}[!htbp]
	\centering
	\includegraphics[width=1.0\linewidth]{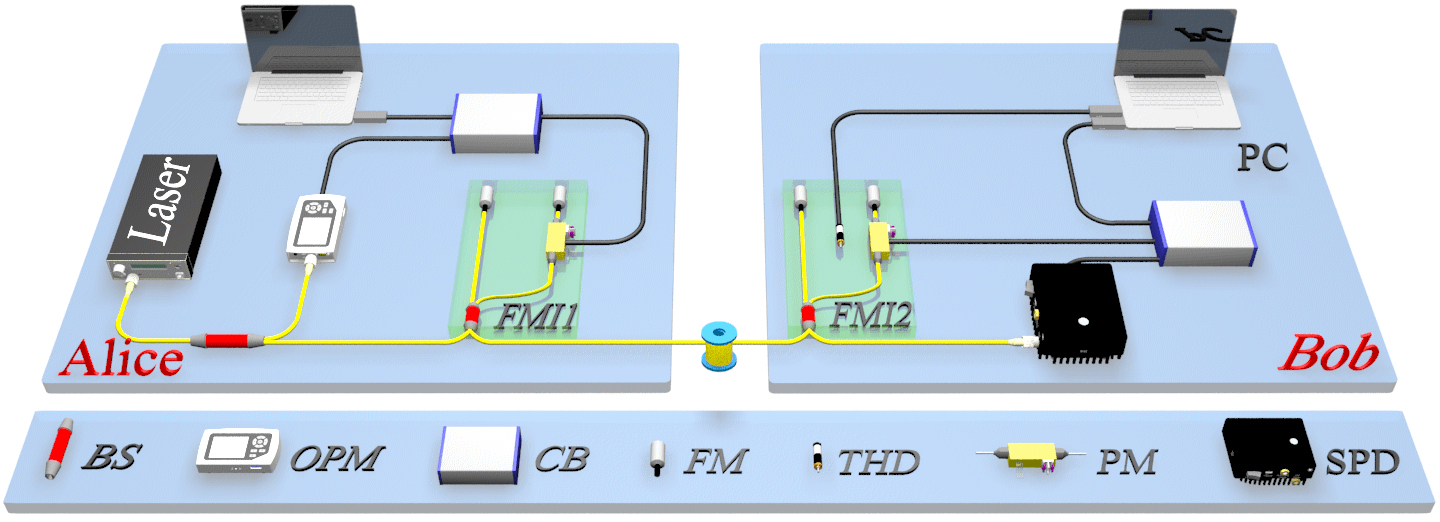}
	\caption{Schematic of the experimental setup. FMI, Faraday-Michelson Interferometer; BS, beam splitter; OPM, optical power meter; CB, control board; FM, Faraday mirror; THD, temperature and humidity detector; PM, phase modulator; SPD, single-photon detector. }
	\label{fig:eta2m2x}
\end{figure}

	The schematic of our experimental setup is displayed in Fig. 5. At Alice's side, a diode laser with a repetition rate of 1 MHz possessing a central wavelength of 1550 nm is sent into a 1:99 beam-splitter (BS) which split the light into two paths, each sent into the optical power meter (OPM) and the FMI respectively. For each light pulse sent into the FMI, it is randomly prepared into one of the BB84 state, encoding information, and sent out to Bob through a commercial standard single-mode optical fiber. At Bob's side, he randomly chooses one of the basis (X or Z) to carry out projection measurements through FMI, and decoding out useful information. For each side, a control board (CB) and a personal computer (PC) are utilized to run the LSTM network and apply proper voltages to the phase modulator (PM). Besides, each of the OPM, the temperature and humidity detector (THD), and the SPD are used to real-time record the laser power, the temperature and humidity, and the counting rate individually. Here the SPD used is a InGaAs single-photon detector working at a gated mode.

\begin{figure}[!htbp]
	\centering
	\includegraphics[width=0.9\linewidth]{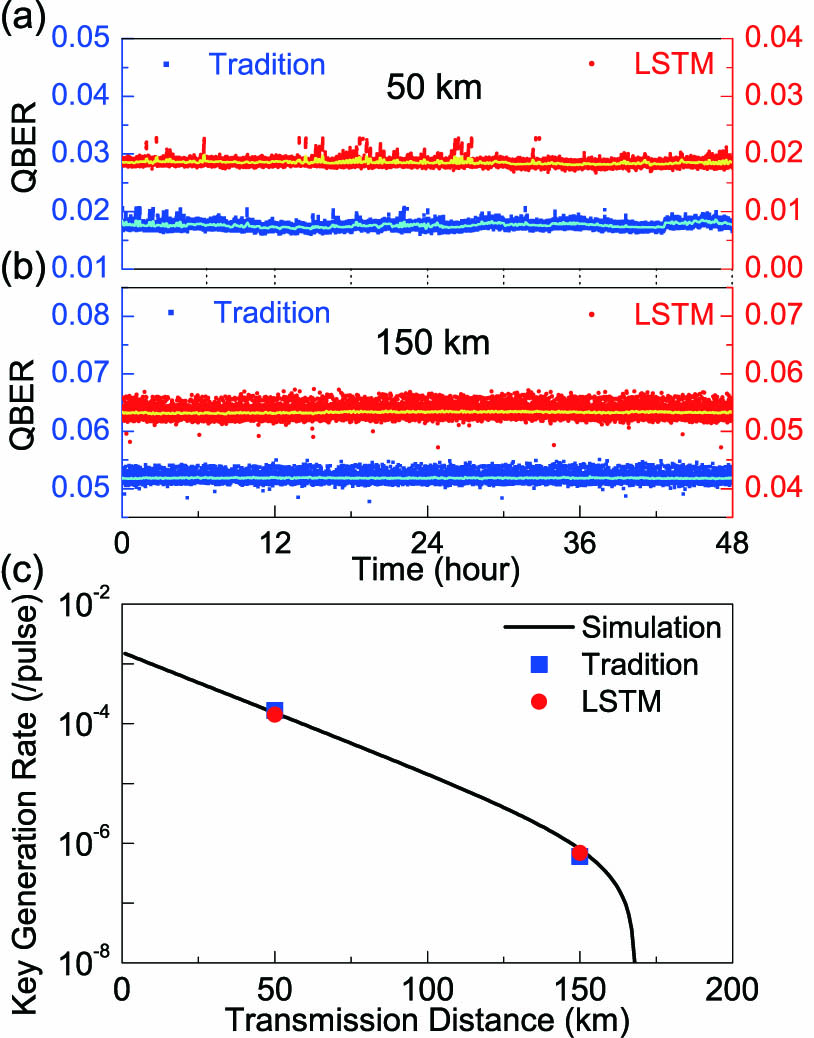}
	\caption{Comparisons between applying traditional "scanning-and-transmitting" program and using present LSTM model based QKD systems. (a) and (b) each refers to the variations of the QBER with time and its average value at the transmission distance of 50 km and 150 km respectively. The QBER of traditional method is linked to left Y-axis and of LSTM model is linked to right Y-axis. (c) The key generation rate versus transmission distance. In each figure, the square points refer to the results of applying traditional "scanning-and-transmitting" program, and the circle points correspond to using our new method.}
	\label{fig:eta2m2x}
\end{figure}

	In our experiment, at each of the transmission distance (50 km and 150 km), we independently run the BB84 QKD system by applying either traditional "scanning-and-transmitting" program  or our present LSTM model for two days, corresponding QBER and counting rates at each time are recorded. During the transmission process, we implement the three-intensity decoy state method, i.e., modulating light pulse into three different intensities ($ \mu $=0.5, $ \nu $=0.1, 0), and calculate the QBER for the above two programs. Corresponding experimental results are shown in Fig. 6. Here the misalignment error rate of the optical system is 0.0123, the detection efficiency and the dark count rate of the single-photon detector is 10\% and $8\times10^{-7}$ respectively. The "duty ratio" of traditional "scanning-and-transmitting" program and our present LSTM model based system is 50\% and 83\%, individually.
  	  
  	Moreover, in order to demonstrate that our LSTM model has the ability of long-time continuous prediction. We run the LSTM model based QKD system for 10 days at the transmission distance of 50 km. Corresponding result is shown in Fig. 7. 
  	  
	\begin{figure}[!htbp]
		\centering
		\includegraphics[width=0.9\linewidth]{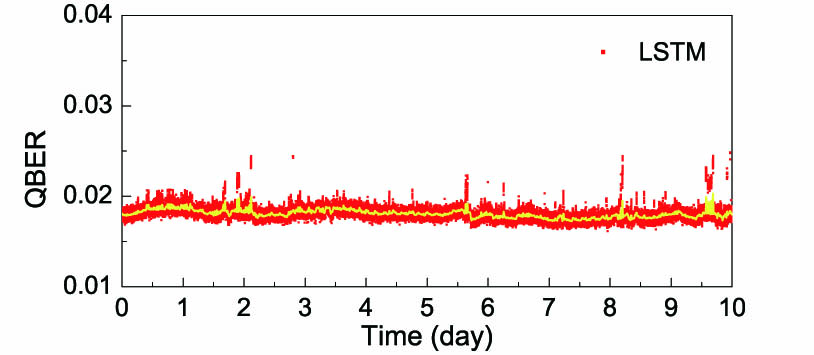}
		\caption{The variations of the QBER with time and its average value within 10 days by running the LSTM model based QKD systems at the transmission distance of 50 km.}
		\label{fig:eta2m2x}
    \end{figure}
	  
\section{ANALYSIS AND DISCUSSION}
In Fig. 6(a) and (b), the variations of the QBER with time and its moving average at the transmission distance of 50 km and 150 km are plotted out individually. In Fig. 6(a), the average QBER of traditional method is 0.0176 and of the LSTM model is 0.0185. In Fig. 6(b), the average QBER of traditional method is 0.0518 and of the LSTM model is 0.532. For simplicity, here we only display the QBER of the signal states $ \mu $, for both methods the QBER value of the decoy state $ \nu $ keeps almost the same level as the signal state at the same transmission distance. In Fig. 6(c), we do comparisons between the theory and the experiment for the quantum key generation rates at different transmission distance, where the line refers to the theoretical predictions, and the points correspond to the experimental data using either traditional "scanning-and-transmitting" program (square points) or our present LSTM model based QKD system (circle points). 

We can see from Fig. 6(a) and (b) that, our LSTM model based QKD system can keep the same level of QBER as traditional "scanning-and-transmitting" program with continuous two days
monitoring, and our present work can obtain similar QBER for each signal pulse as the traditional program. As a result, we can obtain almost the same quantum key generation rate for each transmitted signal pulse as the latter. However, if we take the "duty ratio" into account, our present method has increased the transmission efficiency with at least 33 percent compared with the latter, and this value in principle can be further improved by optimizing the network structure. 

In Fig. 7, the QBER of our LSTM model based QKD system for continuous 10 days measurements has been displayed out, and it shows no trend of deterioration from beginning to end, demonstrating the long-time reliability and stability of our present method.

In both Fig. 6(a) and 7, the QBER appears some rising points. These rising points mainly rooted in some additional disturbances in our laboratory. Then, in the 150 km fiber experiment, we eliminate these additional disturbances, and the QBER results appear no such rising point. 

\section{CONCLUSIONS AND OUTLOOKS}
In conclusion, we have developed a new operating mode for present QKD system called "predicting-and-updating" pattern which can enable the QKD system working reliably and steadily for a long time only with a few scans to update the network. Furthermore, we carry out corresponding experimental demonstration by running a phase-coding BB84 QKD system with either our present method and applying traditional "scanning-and-transmitting" program. Experimental results show that we can keep almost the same level of QBER as the traditional "scanning-and-transmitting" program by applying our present operating mode, but dramatically reduce the scanning and stopping time than the latter, and thus achieve significantly improved transmission efficiency.  

Finally, we should declare that our present work is not limited to phase-coding schemes or BB84 protocols. Here we only take a phase-coding BB84 QKD system as an example to run our model and get excellent experimental results. In principle, this model should also be applicable to any other coding scheme and QKD protocols \cite{MDI,Making,MDIQDS,Overcoming}, since the machine learning model only needs enough training data to construct out corresponding mapping relationship between "features" and "labels", and has no requirements for specific systems or protocols. Therefore, we believe that our present machine learning based operating mode could replace traditional "scanning-and-transmitting" program and play an important role in practical applications of large-scale quantum communication networks in the near future.

\section{ACKNOWLEDGEMENT}
This work has been supported by the National Key Research and Development Program of China (2018YFA0306400, 2017YFA0304100,
2016YFA0302600); National Natural Science Foundation of China (NSFC) (61705110, 11847215, 11774180, 61675235, 61475197, 61590932); China Postdoctoral Science Foundation (2018M642281); Natural Science Foundation of Jiangsu Province (BK20170902); Jiangsu Planned Projects for Postdoctoral Research Funds (2018K185C); University Natural Science Research Project of Jiangsu Province (17KJB510038); The Postgraduate Research \& Practice Innovation Program of Jiangsu Province through Grant No. KYCX19\_0951.

\section*{APPENDIX A: WORKING PRINCIPLE OF LSTM NETWORK}
Considering reasoning about previous events of sequencing events to inform later ones, it’s unclear how a traditional neural network could do. Recurrent neural networks (RNN) \cite{timeSer1} address this issue. They are networks with loops, allowing information to persist and also the natural architecture of neural network to use for time series. However, the standard RNN is unable to solve the long-term Dependencies problem \cite{app1,app2} which means RNNs become unable to learn to connect the previous information as its time gap grows. 

\begin{figure}[!htbp]
	\centering
	\includegraphics[width=0.8\linewidth]{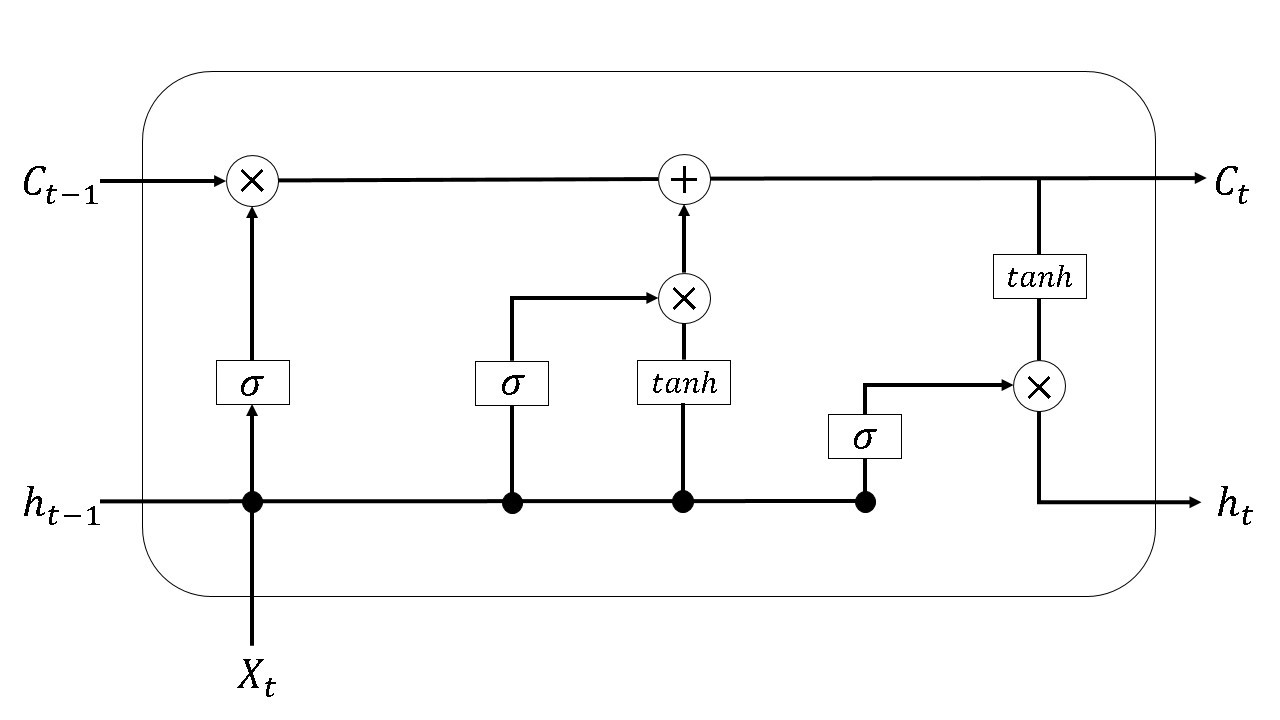}
	\caption{Schematic diagram of the internal structure of a LSTM block.}
	\label{fig:eta2m2x}
\end{figure}

The LSTM is explicitly designed to tackle this problem \cite{app3,app4}. All recurrent neural networks have the form of a chain of repeating modules of neural network. In a standard RNN, this repeating module has a simple structure, usually a single hidden layer. An LSTM also has the chain like structure, but the repeating module has a different structure. Instead of having a single hidden layer, there are four layers, interacting in a very special way. 

The key to LSTMs is the cell state $ C_{t} $ which is kind of like a conveyor belt. Information that has been learned flows along it and runs through the entire chain. The LSTM has the ability to remove or add information to the cell state, regulated by structures called gates. An LSTM block has three of these gates to control the cell state \cite{http}.

Firstly, a sigmoid layer called the "forget gate layer" decides what past information we are going to throw away from the cell state. The input of current moment $ x_{t} $ and the output of last moment $ h_{t-1} $ go through the synapses with certain weights $ W_{f} $ and biases $ b_{f} $ and then are processed by a sigmoid active function $ \sigma $ as following:
	$$
		f_{t} = \sigma(W_{f}\cdot[h_{t-1},x_{t}]+b_{f}).
		\eqno(A1)
	$$
	
The next step is to decide what new information we’re going to store in the cell state. This step consists of two parts. First, also a sigmoid layer called the “input gate layer” decides which values need to be updated. This process is shown in Eq.(A2). Then a tanh layer creates a vector of new candidate values $ \tilde{C_{t}} $ which could be added to the cell. This process is shown in Eq.(A3).
	$$
		i_{t} = \sigma(W_{i}\cdot[h_{t-1},x_{t}]+b_{i}),
		\eqno(A2)
	$$
	$$
		\tilde{C_{t}} = tanh(W_{c}\cdot[h_{t-1},x_{t}]+b_{c}).
		\eqno(A3)
	$$
	
The cell state is updated by multiplying with the two gates mentioned above as shown in Eq.(A4).
	$$
		C_{t} = f_{t}\times C_{t-1}+i_{t}\times\tilde{C_{t}}.
		\eqno(A4)
	$$
	
Finally, the "output gate layer" output the values we need after the filtration of the cell state. We put the cell state through the tanh active function (to push the values to be between -1 and 1) and multiply it by the output of the output gate, so that we only output the parts we decided to. These steps are shown in Eq.(A5) and Eq.(A6).
	$$
		O_{t} = \sigma(W_{o}\cdot[h_{t-1},x_{t}]+b_{o}),
		\eqno(A5)
	$$
	$$
		h_{t} = O_{t}\times tanh(C_{t}).
		\eqno(A6)
	$$

\section*{APPENDIX B: COMPARISONS OF DIFFERENT MODELS}

Before our experiment, we have done some comparisons between the predictions by using different machine learning models, including lower order polynomials, neural network, RNN, and LSTM.  Here we use a representative time series of 3300 data points which its voltage fluctuates violently. The first 70 percent of this dataset is used for training and the second 30 percent for prediction. The inputs of models are temperature T, humidity H, intensity of laser P, and current zero-phase voltage U which is predicted at previous moment. The comparisons of the predicted results from lower order polynomials, a neural network, a RNN, and the LSTM are in Fig. 9. We take the root mean square error (RMSE) of the prediction part as the key index for comparison. Other training details are the same as above. 

\begin{figure}[htbp]
	\centering
	\subfigure{\includegraphics[width=0.49\linewidth]{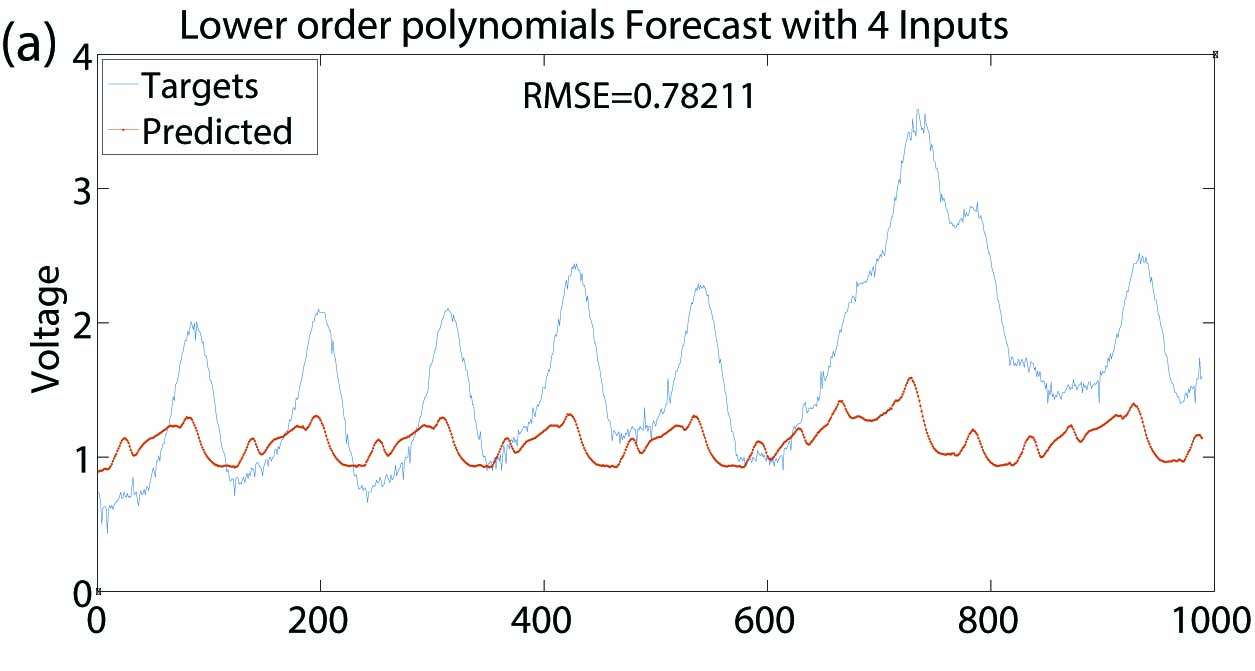}}
	\subfigure{\includegraphics[width=0.49\linewidth]{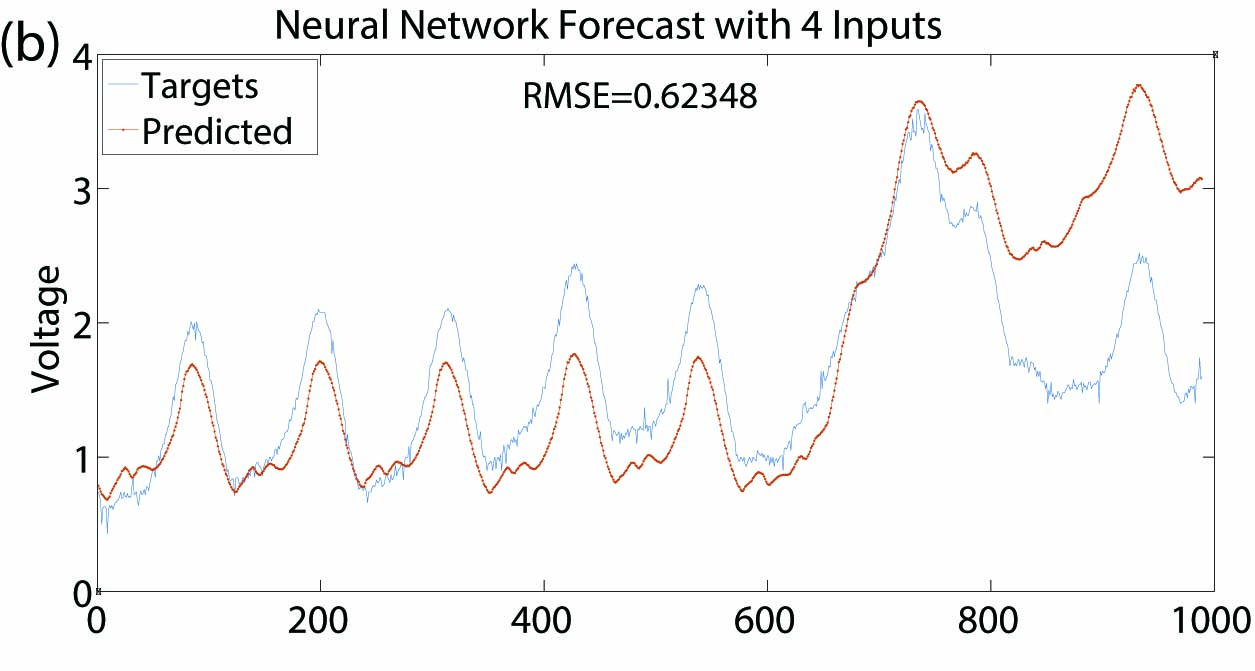}}
	\subfigure{\includegraphics[width=0.49\linewidth]{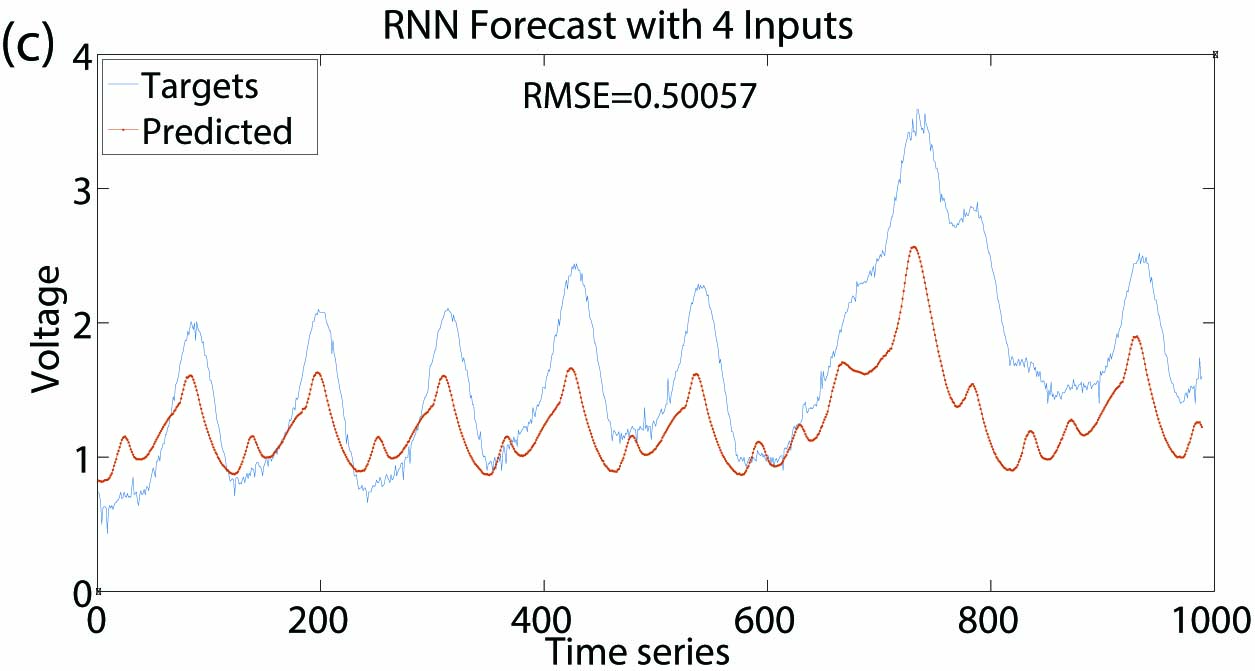}}
	\subfigure{\includegraphics[width=0.49\linewidth]{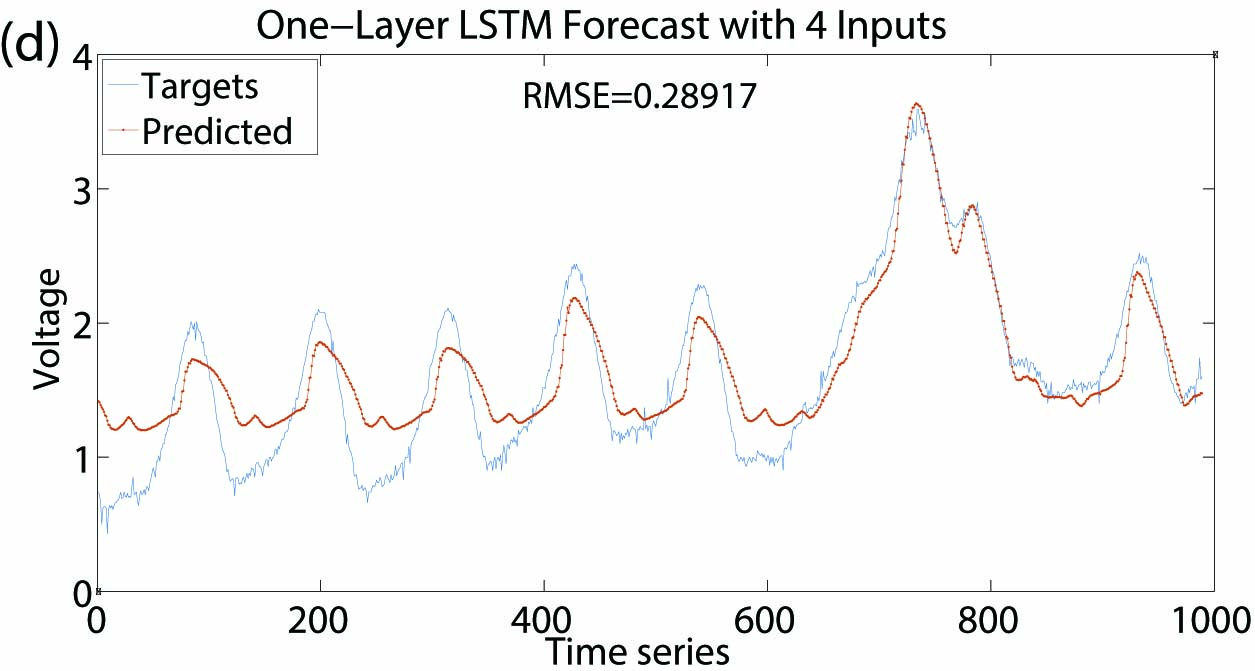}}
	\caption{Comparisons of prediction results from lower order polynomial, neural network, RNN, and LSTM.}
\end{figure}

Compared with RNN and LSTM, the prediction results of neural network and low order polynomial are poor. It can also be seen from FIG. 9(a) and 9(b) that the prediction curve does not fit well with the actual target curve. However, when we compare RNN with LSTM, we can find that LSTM is more accurate than RNN, the RMSE of LSTM is much smaller than that of RNN. Therefore, we find that among the above machine learning models, LSTM model is the best candidate for predicting zero-phase voltages.

\end{document}